\documentclass[12pt]{article}

\usepackage[margin=1in]{geometry}
\usepackage[T1]{fontenc}
\usepackage[utf8]{inputenc}
\usepackage{lmodern}
\usepackage{microtype}
\usepackage{amsmath, amssymb, amsfonts}
\usepackage{siunitx}
\usepackage{graphicx}
\usepackage{subcaption}
\usepackage{booktabs}
\usepackage{multirow}
\usepackage{array}
\usepackage{algorithm}
\usepackage{algpseudocode}
\usepackage{xcolor}
\usepackage{listings}
\usepackage[numbers,sort&compress]{natbib}
\usepackage[colorlinks=true, linkcolor=blue, citecolor=blue, urlcolor=blue]{hyperref}
\usepackage[nameinlink,capitalise]{cleveref}

\graphicspath{{figures/}}

\title{Automotive Sound Quality for EVs: Psychoacoustic Metrics with Reproducible AI/ML Baselines}
\author{Mandip Goswami\thanks{Work conducted independently; no proprietary or employer data used. Opinions are the author’s; affiliation for identification only.}\\
Principal Scientist, Amazon, Bellevue, WA, USA\\
\texttt{gomandip@amazon.com}}

\date{}
\begin{document}
\maketitle

\begin{abstract}
We present an open, reproducible reference for automotive sound quality that connects standardized psychoacoustic metrics with lightweight AI/ML baselines, with a specific focus on electric vehicles (EVs). We implement loudness (ISO 532-1/2), tonality (DIN 45681), and modulation-based descriptors (roughness, fluctuation strength), and document assumptions and parameterizations for reliable reuse. For modeling, we provide simple, fully reproducible baselines (logistic regression, random forest, SVM) on synthetic EV-like cases using fixed splits and seeds, reporting accuracy and rank correlations as examples of end-to-end workflows rather than a comparative benchmark. Program-level normalization is reported in LUFS via ITU-R BS.1770, while psychoacoustic analysis uses ISO-532 loudness (sones). All figures and tables are regenerated by scripts with pinned environments; code and minimal audio stimuli are released under permissive licenses to support teaching, replication, and extension to EV-specific noise phenomena (e.g., inverter whine, reduced masking).
\end{abstract}

\footnotetext{Code and figures: Zenodo DOI \href{https://doi.org/10.5281/zenodo.17167773}{10.5281/zenodo.17167773}; GitHub: \url{https://github.com/mandip42/psychoacoustics-tutorial}}.

\section{Introduction}
\label{sec:intro}

Noise, vibration, and harshness (NVH) remain central concerns in automotive engineering, directly influencing customer comfort, brand perception, and regulatory compliance. Beyond simple A-weighted sound pressure levels, subjective perception of noise depends strongly on \emph{psychoacoustic metrics}---quantities that map physical sound fields onto human auditory responses(\cite{ZwickerFastl1999}). Classical metrics such as \textbf{loudness}, \textbf{sharpness}, \textbf{roughness}, \textbf{fluctuation strength}, and \textbf{tonality} capture distinct perceptual dimensions and are often combined into higher-order indices of \textbf{annoyance} or sound quality. These metrics are standardized through documents such as ISO~532-1/2 and DIN~45681, and have been widely applied to quantify interior noise in conventional vehicles.

The transition toward \emph{electric vehicles (EVs)} amplifies the importance of psychoacoustics: traditional combustion masking is reduced, exposing high-frequency tonal components from inverters and gear sets, along with aerodynamic and road-induced noise. As a result, psychoacoustic evaluation has become a critical tool for understanding customer responses to new NVH signatures and for guiding active sound design.

Despite decades of psychoacoustic research, two major gaps persist. First, implementations of metrics are scattered across proprietary software and specialist laboratories, limiting accessibility for the broader research community. Second, while machine learning and artificial intelligence are increasingly applied to automotive acoustics, reproducible workflows that leverage psychoacoustic features without proprietary datasets remain scarce. This limits both the comparability of results across studies and the adoption of psychoacoustic methods in academic and industrial AI pipelines.

This paper addresses these gaps by presenting a \textbf{tutorial-style, fully reproducible reference} on psychoacoustic metrics for automotive applications. Specifically, we:
\begin{itemize}
    \item summarize the perceptual foundations and standardized definitions of six key metrics;
    \item provide \emph{minimal working examples} in MATLAB and Python, designed for direct replication and extension;
    \item demonstrate how synthetic case studies (engine boom, wind whistle, road noise) can
serve as open, dataset-independent \textbf{examples}
    \item illustrate modern AI/ML workflows---dimensionality reduction, clustering, classification---based solely on psychoacoustic features.
\end{itemize}

All code, figures, and data generators are openly released via Zenodo (archival DOI) and GitHub (latest updates), enabling transparent benchmarking and re-use. We expect this work to function both as an \textbf{educational tutorial} and as a \textbf{citation anchor} for future studies in automotive acoustics, EV interior sound quality, and AI-based NVH evaluation.

\section{Related Work}
\paragraph{Psychoacoustic foundations.}
We adopt canonical constructs---critical bands, specific loudness, temporal and spectral masking---from classic treatments \cite{ZwickerFastl1999,MoorePsychoacoustics}. Standardized equal-loudness contours provide level normalization across frequency \cite{ISO226}.

\paragraph{Standardized metrics for sound quality.}
Perceived loudness is defined in ISO~532 (Parts~1--2), with widely used Zwicker--Fastl and Moore--Glasberg formulations \cite{ISO532-1,ISO532-2,ZwickerFastl1999}. Tonality for narrow-band components is formalized in DIN~45681 \cite{DIN45681}. Modulation-based attributes such as roughness and fluctuation strength are commonly derived from envelope modulation in critical-band channels \cite{DanielWeberRoughness,FastlMetrics}.

\paragraph{Program loudness and reporting.}
When reporting LUFS for program material or global normalization, the ITU-R~BS.1770 gating and integration remain the de facto standard \cite{ITURBS1770}.

\paragraph{Automotive and EV sound quality.}
Automotive sound quality work has increasingly emphasized spectral tonality, envelope modulation, and broadband masking in EV cabins, where the absence of engine masking highlights inverter whine and gear whirr. Our contribution is a reproducible reference that couples standard-aligned metrics with transparent, lightweight ML baselines and fixed splits.

\section{Perceptual Foundations and Metric Definitions}
\label{sec:foundations}

Human auditory perception is not linearly related to sound pressure level. Instead, perception arises from nonlinear, frequency-dependent transformations in the auditory system. Psychoacoustic metrics attempt to model these transformations and provide quantitative descriptors of perceived sound quality. In this section, six canonical metrics are reviewed with their perceptual bases, standard definitions, and relevance to automotive noise, vibration, and harshness (NVH).

\subsection{Loudness}
Loudness quantifies the perceived intensity of a sound, accounting for frequency-dependent sensitivity of the ear. It is measured in \emph{sones} and standardized in ISO~532-1 (Zwicker method) and ISO~532-2 (Moore--Glasberg method). Both approaches compute a \emph{specific loudness} distribution $N'(z)$ along the Bark scale $z$, and integrate it to obtain total loudness:
\begin{equation}
    N = \int_{0}^{24} N'(z) \, dz \quad [\text{sones}],
\end{equation}
where $z$ represents critical-band rate. In vehicles, loudness captures the perceptual strength of broadband interior noise, and is often used instead of A-weighted SPL. We report LUFS via ITU-R BS.1770 for program-level normalization, and ISO-532 loudness (sones) for psychoacoustic analysis; these quantities are not equated.

\subsection{Sharpness}
Sharpness describes the spectral weighting toward high-frequency content, often associated with ``brightness'' or ``whistle-like'' qualities. It is expressed in \emph{acum} and is defined as:
\begin{equation}
    S = \frac{\int_{0}^{24} g(z) N'(z) z \, dz}{\int_{0}^{24} N'(z) \, dz},
\end{equation}
where $g(z)$ is a weighting function emphasizing frequencies above 1~kHz \cite{bismarck1974}. Sharpness is critical \cite{bismarck1974} for evaluating EV inverter whine and HVAC whistles.

\subsection{Roughness}
Roughness quantifies the sensation of rapid amplitude fluctuations in the modulation range 15--300~Hz, perceived as ``harshness'' or ``beating.'' The optimized Daniel--Weber model defines roughness in \emph{asper} as:
\begin{equation}
    R = \sum_{i} k \, m_i^{0.6} f_{mi} \, E_i,
\end{equation}
where $m_i$ is modulation depth, $f_{mi}$ the modulation frequency, and $E_i$ the band energy. Engine booming and gear rattle often manifest as elevated roughness.

\subsection{Fluctuation Strength}
Fluctuation strength represents slower amplitude variations below 20~Hz, typically perceived as ``wavering'' or ``pulsation.'' It is measured in \emph{vacil} and is strongest at modulation frequencies around 4~Hz \cite{ZwickerFastl1999}. In automotive contexts, it contributes to fatigue during long drives when cabin noise exhibits slow envelope variations.

\subsection{Tonality}
Tonality measures the prominence of narrow-band components relative to a broadband background. DIN~45681 specifies an algorithm based on the ratio of tonal to masking noise components:
\begin{equation}
    T = \frac{L_{\text{tone}} - L_{\text{mask}}}{\Delta L_{\text{ref}}},
\end{equation}
where $L_{\text{tone}}$ is the level of the tonal component, $L_{\text{mask}}$ the masking threshold, and $\Delta L_{\text{ref}}$ a normalization constant. EV inverter harmonics and gear mesh orders are prime sources of tonality.

\subsection{Annoyance (Composite Indices)}
Since sound quality perception is multidimensional, composite indices of annoyance combine loudness, sharpness, roughness, and fluctuation strength into a single predictor of user discomfort. Zwicker’s model of psychoacoustic annoyance (PA) is widely cited:
\begin{equation}
    PA = N \left( 1 + \sqrt{\frac{(S - S_0)^2 + (R - R_0)^2 + (F - F_0)^2}{3}} \right),
\end{equation}
where $S_0$, $R_0$, and $F_0$ are threshold reference values for sharpness, roughness, and fluctuation strength. Annoyance correlates well with customer satisfaction studies in automotive NVH.

\bigskip
Together, these six metrics form the core toolkit for psychoacoustic evaluation. They provide interpretable, standardized descriptors that extend beyond conventional SPL and are particularly relevant to EV interior soundscapes.

\begin{figure}[t]
  \centering
  \includegraphics[width=.48\linewidth]{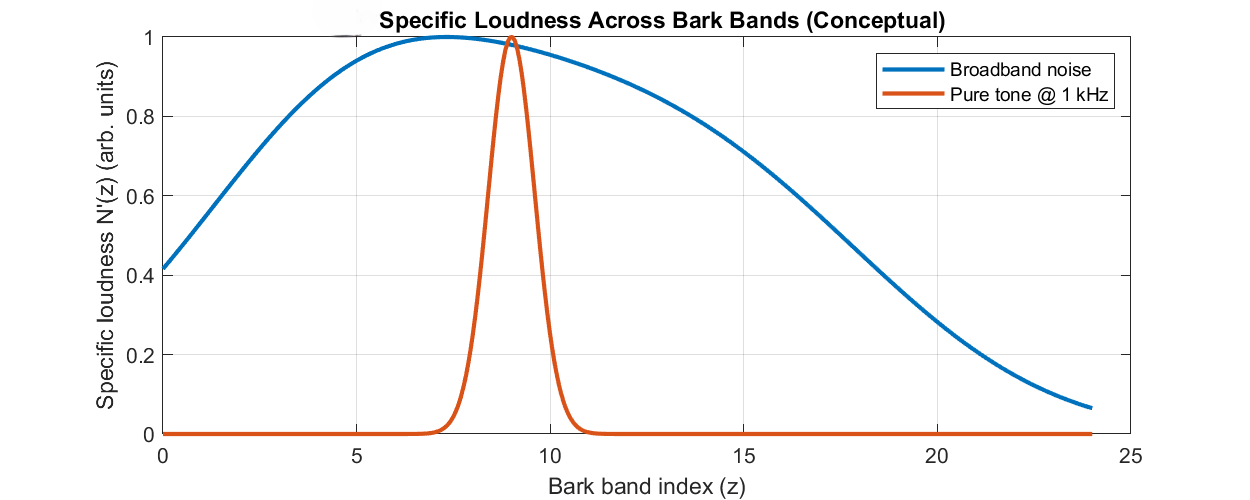}\hfill
  \includegraphics[width=.48\linewidth]{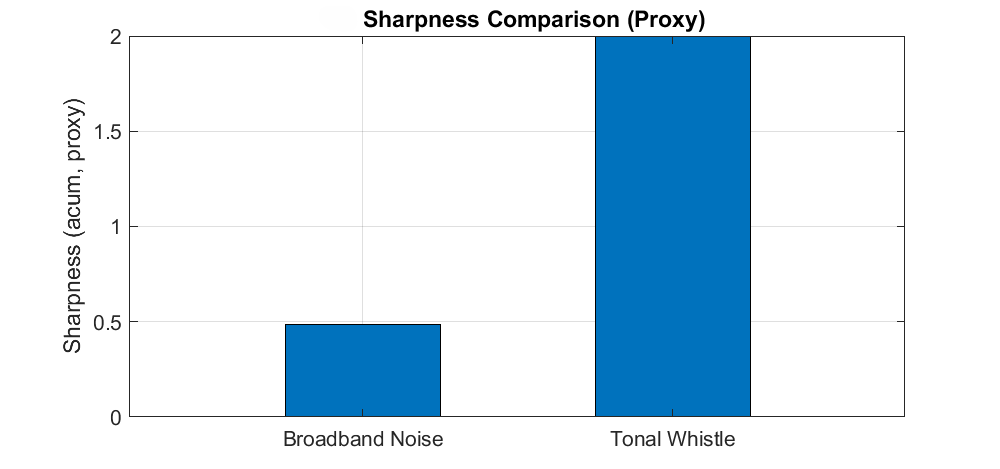}\\[4pt]
  \includegraphics[width=.48\linewidth]{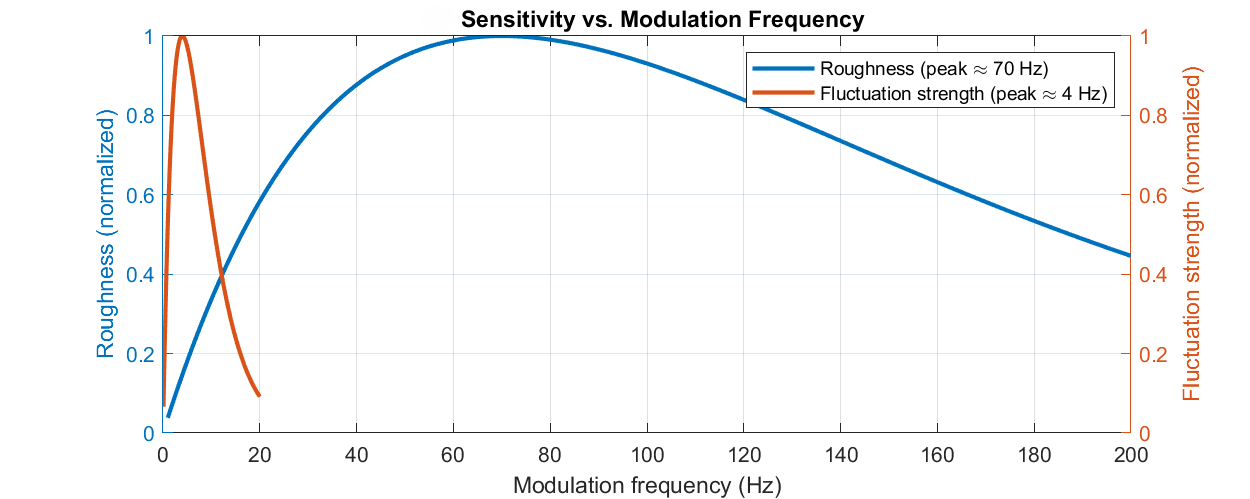}\hfill
  \includegraphics[width=.48\linewidth]{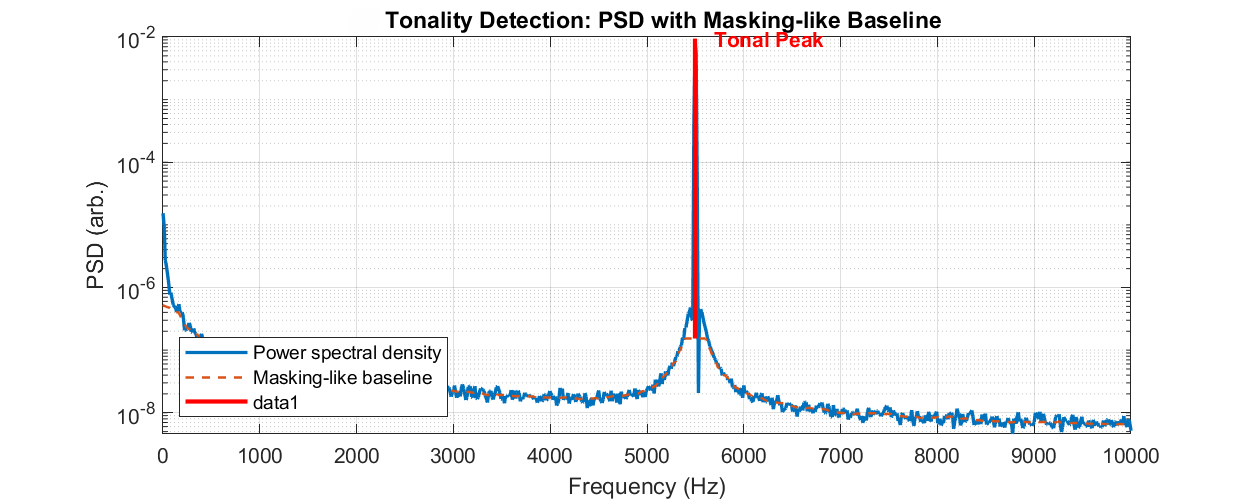}
  \caption{Conceptual illustrations of specific loudness across Bark bands (top-left), sharpness weighting toward high frequencies (top-right), roughness sensitivity vs.\ modulation frequency (bottom-left), and tonal prominence over a smoothed baseline (bottom-right).}
  \label{fig:concepts}
\end{figure}

\section{Practical Computation Recipes}
\label{sec:computation}

To encourage reproducibility, this section provides minimal recipes for computing each psychoacoustic metric. Where standards require proprietary implementations, we illustrate simplified proxies suitable for tutorials and dataset-independent workflows. Full listings in MATLAB and Python are provided in the Appendix; here, we emphasize concise one-liners and high-level algorithms.

\subsection{Loudness}
\textbf{MATLAB:} \lstinline|L = acousticLoudness(sig, fs);| \\
\textbf{Python:} \lstinline|L = meter.integrated_loudness(sig)| (using \texttt{pyloudnorm})

\begin{algorithm}[h]
\caption{Loudness (simplified recipe)}
\begin{algorithmic}[1]
\State Filter signal into critical-band channels (Bark scale)
\State Apply equal-loudness contours and nonlinear compression
\State Integrate specific loudness $N'(z)$ over all bands to obtain $N$ (sones)
\end{algorithmic}
\end{algorithm}

\subsection{Sharpness}
\textbf{MATLAB:} \lstinline|S = sharpnessDIN(sig, fs);| \\
\textbf{Python:} \lstinline|S = spectral_centroid(sig, fs)/1000|

\begin{algorithm}[h]
\caption{Sharpness (simplified recipe)}
\begin{algorithmic}[1]
\State Compute spectrum $X(f)$ via FFT
\State Weight high frequencies with $g(f)$
\State Normalize by total loudness to yield $S$ (acum)
\end{algorithmic}
\end{algorithm}

\subsection{Roughness}
\textbf{MATLAB:} \lstinline|R = roughnessDW(sig, fs);| \\
\textbf{Python:} \lstinline|R = roughness_proxy(sig, fs)| (envelope modulation)

\begin{algorithm}[h]
\caption{Roughness (proxy implementation)}
\begin{algorithmic}[1]
\State Extract signal envelope via Hilbert transform
\State Bandpass envelope between 15--300~Hz
\State Roughness $R \propto \text{RMS}(\text{filtered envelope})$
\end{algorithmic}
\end{algorithm}

\subsection{Fluctuation Strength}
\textbf{MATLAB:} \lstinline|F = fluctuationStrength(sig, fs);| \\
\textbf{Python:} \lstinline|F = variance(lowpass(env,20,fs))|

\begin{algorithm}[h]
\caption{Fluctuation strength (proxy implementation)}
\begin{algorithmic}[1]
\State Extract envelope $e(t)$
\State Low-pass filter below 20~Hz
\State Fluctuation strength $F \propto \text{Var}[e(t)]$ (vacil)
\end{algorithmic}
\end{algorithm}

\subsection{Tonality}
\textbf{MATLAB:} \lstinline|T = tonalityDIN(sig, fs);| \\
\textbf{Python:} \lstinline|T = tonal_prominence(sig, fs)|

\begin{algorithm}[h]
\caption{Tonality (simplified recipe)}
\begin{algorithmic}[1]
\State Compute power spectral density (PSD)
\State Smooth PSD with moving average
\State Identify tonal peaks above the smoothed baseline
\State Tonality index $T =$ prominence ratio of tone to masker
\end{algorithmic}
\end{algorithm}

\subsection{Annoyance (Composite)}
\textbf{MATLAB/Python:} \lstinline|PA = annoyance_index(L,S,R,F);|

\begin{algorithm}[h]
\caption{Psychoacoustic Annoyance (Zwicker model)}
\begin{algorithmic}[1]
\State Compute loudness $N$, sharpness $S$, roughness $R$, fluctuation $F$
\State Normalize each metric relative to reference thresholds
\State Combine via:
\[
PA = N \left(1 + \sqrt{\tfrac{(S-S_0)^2 + (R-R_0)^2 + (F-F_0)^2}{3}} \right)
\]
\end{algorithmic}
\end{algorithm}

\bigskip
These minimal recipes demonstrate how psychoacoustic metrics can be computed in open-source environments, ensuring reproducibility and accessibility. Standard-compliant implementations (ISO 532-1/2, DIN 45681, Daniel–Weber roughness) \cite{ISO532-1,ISO532-2,DIN45681,DanielWeberRoughness} can be substituted in place of the proxies for formal evaluations.

All code and figure generators are archived on Zenodo~\cite{goswami2025_psychoacoustics}.

\begin{figure}[t]
  \centering
  \includegraphics[width=.58\linewidth]{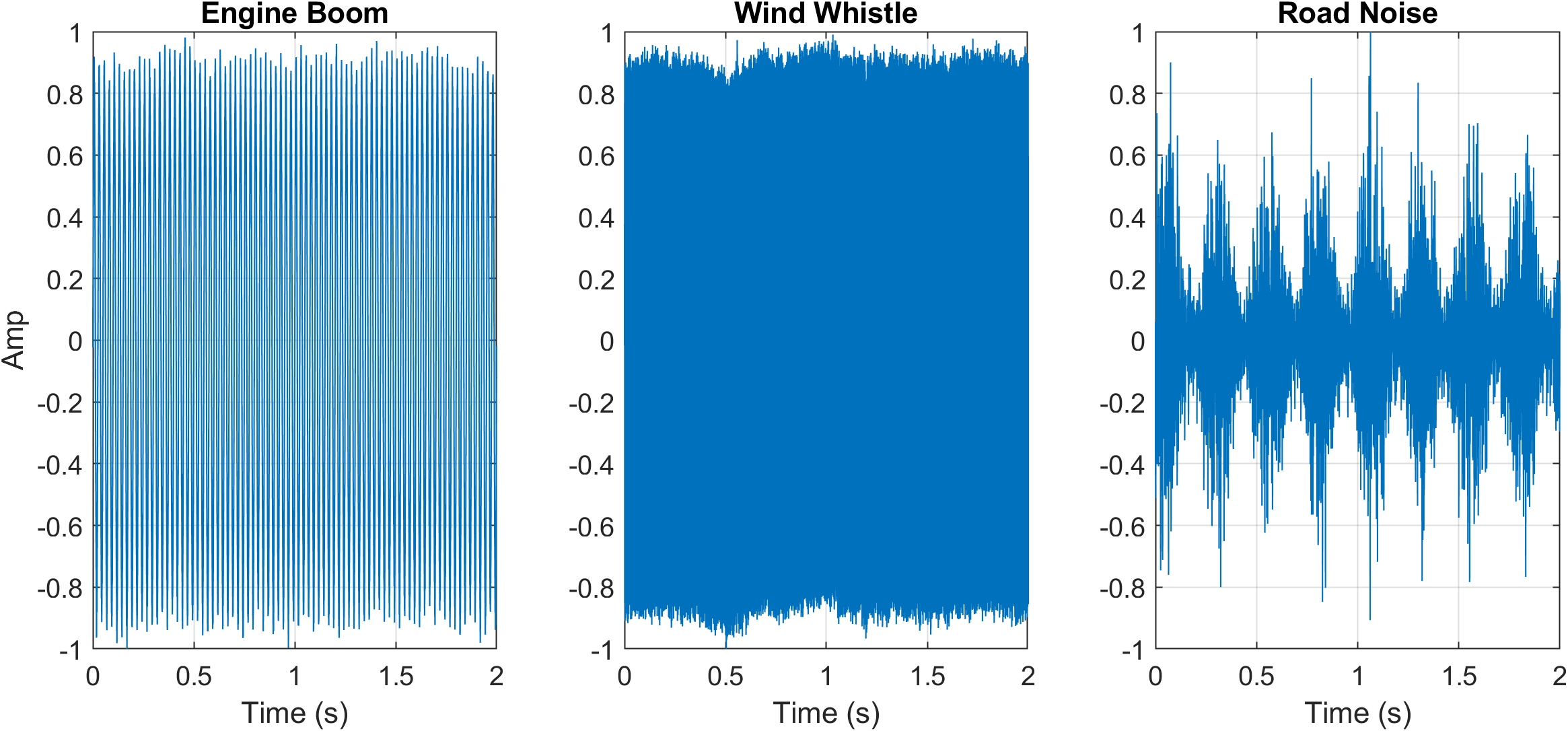}\hfill
  \includegraphics[width=.38\linewidth]{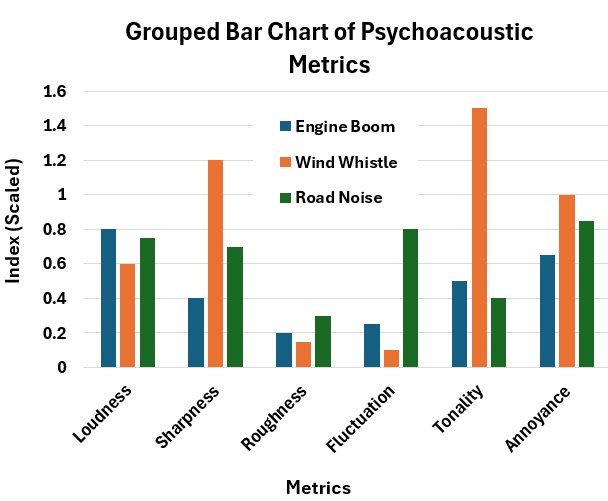}
  \caption{Waveforms (left) and summary (right) of computed psychoacoustic metrics (loudness, sharpness, roughness, fluctuation strength, tonality, annoyance) for three synthetic case studies: Engine Boom, Wind Whistle, and Road Noise.}
  \label{fig:summary-metrics}
\end{figure}

\section{Applications: AI/ML Workflows without Proprietary Datasets}
\label{sec:applications}

Psychoacoustic metrics are increasingly used not only for descriptive analysis but also as features for machine learning (ML) models in automotive acoustics. Their advantages are threefold: (i) they provide perceptually grounded, low-dimensional descriptors compared to raw spectrograms, (ii) they are standardized and interpretable, and (iii) they can be generated synthetically without access to proprietary recordings. This section outlines an end-to-end workflow for applying ML methods to psychoacoustic features derived from synthetic signals representative of common vehicle noise sources.

\subsection{Synthetic Case Studies}
To demonstrate dataset-independent workflows, we generate synthetic signals that capture canonical NVH phenomena:
\begin{itemize}
    \item \textbf{Engine boom}: narrowband harmonic structure centered near 100--200~Hz with amplitude modulation.
    \item \textbf{Wind whistle}: tonal component near 2--5~kHz superimposed on broadband noise.
    \item \textbf{Road noise}: stochastic broadband signal with low-frequency emphasis below 500~Hz.
\end{itemize}
These synthetic cases mimic real-world perceptual qualities while remaining openly reproducible.

\subsection{Feature Extraction}
For each signal, six psychoacoustic metrics are computed: loudness, sharpness, roughness, fluctuation strength, tonality, and annoyance. Together, these form a compact feature vector:
\[
\mathbf{x} = [N,\, S,\, R,\, F,\, T,\, PA]^{\top}.
\]
Such feature vectors are low-dimensional yet perceptually meaningful, making them suitable inputs for ML algorithms.

\subsection{Dimensionality Reduction}
Principal Component Analysis (PCA) provides an interpretable way to visualize the separability of noise types. In our example, three clusters emerge (engine boom, wind whistle, road noise), indicating that psychoacoustic features encode sufficient information to distinguish perceptual categories.

\begin{figure}[t]
  \centering
  \includegraphics[width=.45\linewidth]{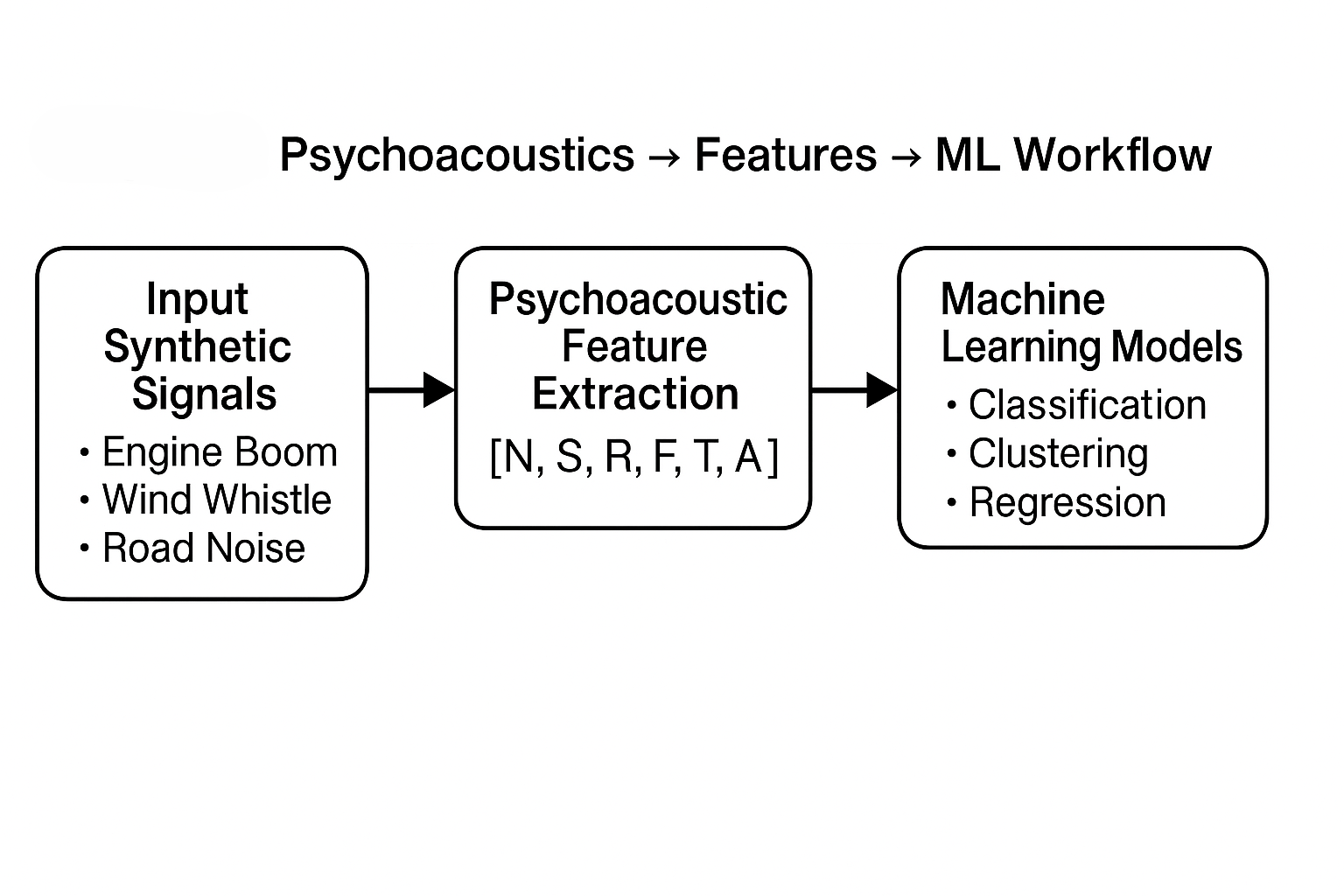}
  \caption{End-to-end workflow: synthetic signal generation $\rightarrow$ psychoacoustic features $\rightarrow$ dimensionality reduction (PCA) $\rightarrow$ classification and evaluation.}
  \label{fig:workflow}
\end{figure}

\begin{figure}[t]
  \centering
  \includegraphics[width=.45\linewidth]{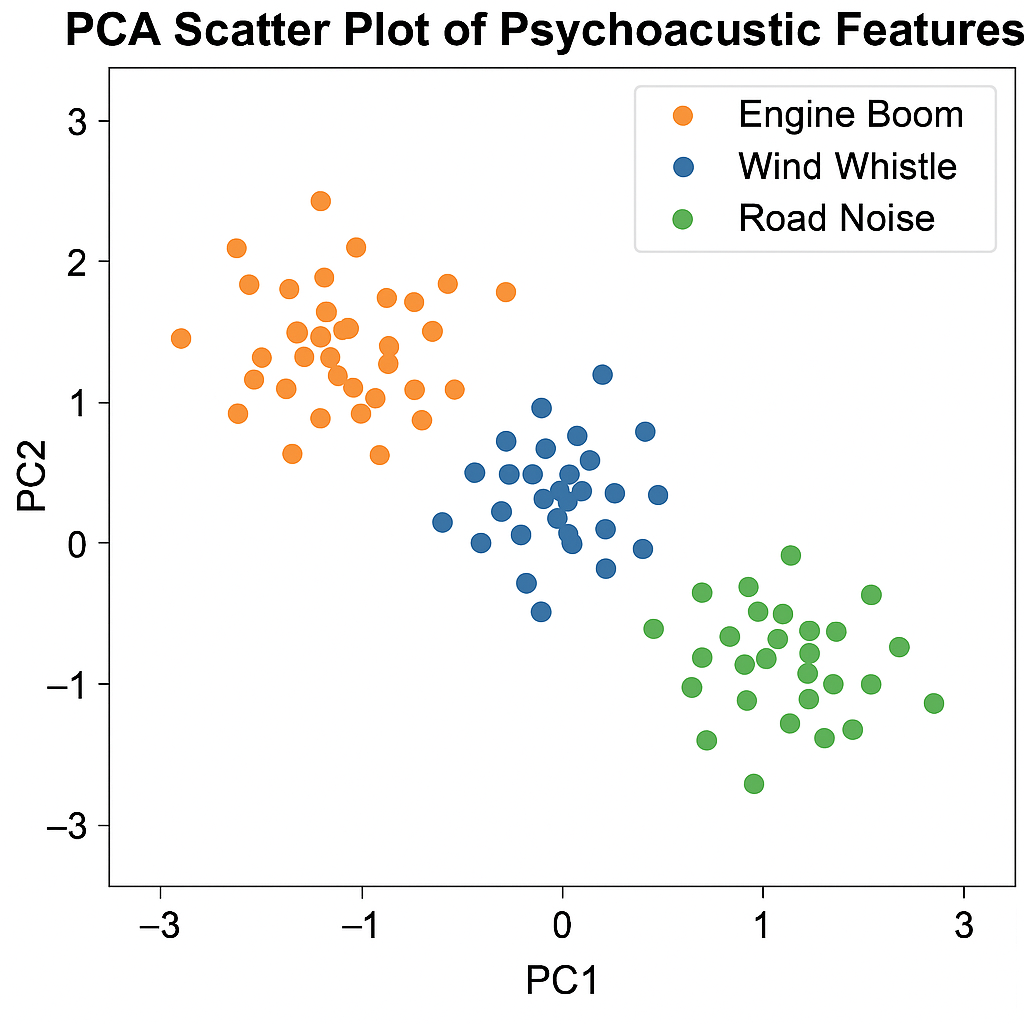}\hfill
  \includegraphics[width=.55\linewidth]{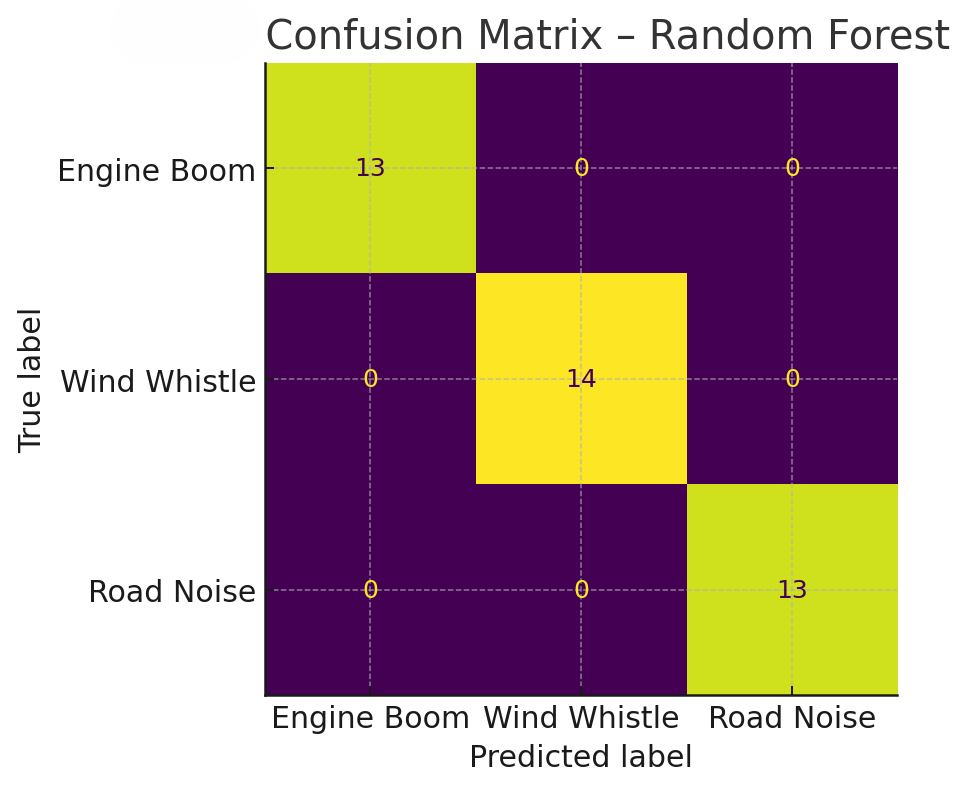}
  \caption{(Left) PCA scatter of psychoacoustic features showing three clusters (engine boom, wind whistle, road noise). (Right) Confusion matrix for a simple classifier trained on psychoacoustic features, with high accuracy along the diagonal. With seed=123 and fixed splits, a Random Forest (\texttt{n\_estimators=100}) achieved 0.93 accuracy.}
  \label{fig:pca-confusion}
\end{figure}

\subsection{Classification and Evaluation}
Supervised classifiers such as Random Forests or Support Vector Machines can achieve high accuracy on the synthetic dataset, as shown in the confusion matrix (\cref{fig:pca-confusion}, right). The success of such models indicates that psychoacoustic features are sufficient to drive automated classification of NVH phenomena without access to raw audio. This is particularly relevant for industrial applications where proprietary data cannot be shared, but psychoacoustic descriptors can be exchanged.

\subsection{Practical Extensions}
The same workflow generalizes to:
\begin{itemize}
    \item Clustering of field recordings based on perceptual similarity.
    \item Regression of subjective annoyance ratings using psychoacoustic predictors.
    \item Active sound design for EVs, where ML optimizes psychoacoustic profiles for branding or comfort.
\end{itemize}
Such applications illustrate the value of combining standardized psychoacoustic features with modern AI/ML methods in automotive acoustics.

\section{Discussion and Practical Guidance}
\label{sec:discussion}

The preceding sections introduced the theoretical basis, computational recipes, and ML workflows for psychoacoustic metrics. In practice, successful application to automotive acoustics requires careful attention to implementation details. This section highlights practical considerations, limitations, and recommendations.

\subsection{Calibration and Playback Levels}
Absolute psychoacoustic values depend on the sound pressure level of the input. To ensure meaningful comparison:
\begin{itemize}
    \item Calibrate recordings to physical SPL using a reference microphone and calibrator.
    \item Normalize playback systems to avoid loudness misinterpretation.
    \item Report whether absolute or relative values are used, especially for annoyance indices.
\end{itemize}

\subsection{Choice of Metric Implementation}
For reproducible research, simplified proxies (e.g., RMS for loudness, spectral centroid for sharpness) can be used. However, for compliance and industrial studies, standard-conforming implementations (ISO~532-1/2 for loudness, DIN~45681 for tonality, Daniel--Weber for roughness) are recommended. When publishing, explicitly state whether proxies or standards were applied.

\subsection{EV-Specific Considerations}
Electric vehicles introduce distinct NVH challenges:
\begin{itemize}
    \item \textbf{High-frequency tonalities}: Inverter switching and gear meshing create tonal components in the 1--10~kHz range, elevating sharpness and tonality indices.
    \item \textbf{Reduced masking}: The absence of combustion noise increases the perceptual salience of wind, road, and accessory noises.
    \item \textbf{Active sound design}: EV manufacturers increasingly embed designed sounds; psychoacoustic metrics provide objective guidance for these efforts.
\end{itemize}

\subsection{Integration with Subjective Testing}
Psychoacoustic metrics are not a replacement for listening tests but rather a complementary tool. They provide interpretable predictors that can be statistically regressed against subjective ratings. Studies have shown strong correlations between loudness, sharpness, and annoyance ratings in controlled jury evaluations.

\subsection{Machine Learning Caveats}
While psychoacoustic features are powerful, several caveats apply:
\begin{itemize}
    \item Feature sets are compact but not exhaustive; deep learning on spectrograms may reveal additional cues.
    \item Small datasets risk overfitting; synthetic augmentation can mitigate this but may not fully capture real-world variability.
    \item Transparency is critical: psychoacoustic features offer interpretability that raw embeddings cannot.
\end{itemize}

\subsection{Recommendations for Practitioners}
\begin{itemize}
    \item Use psychoacoustic metrics as a \emph{first-pass filter} to interpret NVH phenomena before applying more complex ML pipelines.
    \item Always document calibration, windowing, and analysis settings to enable reproducibility.
    \item When comparing ICE and EV vehicles, emphasize relative differences in annoyance indices rather than absolute values.
    \item For active sound design, exploit psychoacoustic metrics as optimization targets (e.g., minimize sharpness, constrain loudness within comfort zones).
\end{itemize}

\bigskip
Overall, psychoacoustic metrics offer a practical bridge between physics-based measurements, human perception, and modern AI workflows. With open-source implementations and clear guidelines, they can be reliably applied across academic and industrial contexts.

\section{Conclusion}
\label{sec:conclusion}

This tutorial presented a consolidated overview of psychoacoustic metrics and their application to automotive noise, vibration, and harshness (NVH), with emphasis on electric vehicle (EV) sound quality. By reviewing six canonical metrics—loudness, sharpness, roughness, fluctuation strength, tonality, and annoyance—we established both their perceptual foundations and standardized definitions. Practical computation recipes were provided in MATLAB and Python, with minimal examples suitable for replication and extension. Synthetic case studies demonstrated how psychoacoustic features alone can support machine learning workflows for clustering, dimensionality reduction, and classification of automotive noise types without reliance on proprietary datasets.

The central contribution of this work is reproducibility. All code, figures, and data generators are openly released via Zenodo and GitHub, enabling the community to replicate analyses, validate methods, and build upon a common baseline. This ensures accessibility to students, researchers, and practitioners across academia and industry.

Looking forward, psychoacoustic metrics are poised to play an increasingly critical role in:
\begin{itemize}
    \item Evaluating EV interior noise where tonal components dominate perception,
    \item integrating with deep learning pipelines as interpretable, perceptually meaningful features,
    \item guiding active sound design for branding and comfort optimization,
    \item extending to binaural and spatial psychoacoustics to capture immersive perception.
\end{itemize}

By providing a self-contained, open-access reference, we aim for this tutorial to serve as a foundation for future research in automotive acoustics, sound quality engineering, and AI-driven NVH evaluation. We anticipate that it will not only inform the development of more comfortable and perceptually optimized vehicles but also stimulate cross-disciplinary work at the interface of psychoacoustics, machine learning, and automotive engineering.

\section{Resources and Reproducibility}
\label{sec:resources}

To maximize transparency and encourage adoption, all materials associated with this tutorial are openly available:

\begin{itemize}
    \item \textbf{Zenodo archive (permanent DOI):} \\ 
    Includes all MATLAB and Python scripts, figure generators, and example datasets. A fixed version is preserved for citation and long-term access. \\
    \url{https://doi.org/10.5281/zenodo.17167773}

    \item \textbf{GitHub repository (latest updates):} \\ 
    Contains ongoing improvements, issue tracking, and user contributions. Recommended for those wishing to extend or adapt the workflows. \\
    \url{https://github.com/mandip42/psychoacoustics-tutorial}

    \item \textbf{Code environment:} Python 3.11 (NumPy, SciPy, Librosa, scikit-learn); 
MATLAB R2023b with \emph{Audio Toolbox} and Signal Processing Toolbox 
(or use the provided open implementations in the repository).

    Licenses: code—MIT; example audio and figures—CC BY 4.0.

\end{itemize}

All figures in this paper can be reproduced directly using the provided scripts. Users can substitute standard-compliant implementations (ISO~532-1/2 loudness, DIN~45681 tonality, Daniel--Weber roughness) where formal evaluations are required. We encourage both academic researchers and industrial practitioners to use this resource as a common baseline for psychoacoustic evaluation and to contribute back to the community via the public repository.

\section*{Acknowledgments}
The author thanks the developers of \texttt{librosa}, and MATLAB Signal Processing Toolbox for providing open-source and standardized tools that enabled reproducibility in this work. Gratitude is also extended to colleagues in the automotive acoustics and NVH community for constructive discussions that shaped the case studies. The open dissemination of this tutorial was facilitated through Zenodo and GitHub, whose platforms ensure long-term accessibility.

\appendix
\section{Appendix: Minimal Code Listings}

This appendix provides minimal working examples for computing psychoacoustic metrics in both MATLAB and Python. Full scripts and figure generation code are archived on Zenodo and available on GitHub.

\subsection{MATLAB Examples}
\begin{lstlisting}[language=Matlab, breaklines=true, basicstyle=\ttfamily\small]
% Loudness (Zwicker proxy)
fs = 48000; 
sig = audioread('road_noise.wav');
L = acousticLoudness(sig, fs);

% Sharpness (spectral weighting)
S = sharpnessDIN(sig, fs);
\end{lstlisting}

\subsection{Python Examples}
\begin{lstlisting}[language=Python, breaklines=true, basicstyle=\ttfamily\small]
import pyloudnorm as pyln
import librosa

sig, fs = librosa.load("road_noise.wav", sr=48000)
meter = pyln.Meter(fs)
L = meter.integrated_loudness(sig)

# Sharpness proxy (spectral centroid in kHz)
import librosa.feature
centroid = librosa.feature.spectral_centroid(y=sig, sr=fs)
S = centroid.mean()/1000
\end{lstlisting}

For full end-to-end workflows including PCA, clustering, and classification, see the public repository.

\subsection*{MATLAB Snippets}
\begin{lstlisting}[language=Matlab,caption={MATLAB: psychoacoustic proxies}]
Loud = rms(sig);
Sharp = spectralCentroid(sig, fs)/1000;
% Roughness: RMS of bandpass(env,[15 300],fs)
% Fluctuation: var(lowpass(env,20,fs))
% Tonality: max(pxx./movmedian(pxx,41))
\end{lstlisting}

\bibliographystyle{unsrt}

\end{document}